\documentstyle[twocolumn,prl,aps,epsf,floats]{revtex} 
\begin{document} 
\title{Charge-Density-Wave Induced Modifications to the Quasiparticle Self-
energy in 2H-TaSe$_2$} 
\author{T. Valla$^1$, A. V. Fedorov$^1$, P. D. Johnson$^1$, J. Xue$^2$, K. E. 
Smith$^2$ and F. J. DiSalvo$^3$}
\address{$^1$ Department of Physics, Brookhaven National Laboratory, Upton, NY, 
11973-5000}
\address{$^2$ Department of Physics, Boston University, 590 Commonwealth Av, 
Boston, MA 02215}
\address{$^3$ Department of Chemistry, Baker Laboratory, Cornell University, 
Ithaca, New York 14853}

\address{ {\em \bigskip \begin{quote}
The self-energy of the photo-hole in 2H-TaSe$_2$ is measured by angle-resolved 
photoemission spectroscopy (ARPES) as a function of binding energy and 
temperature. In the charge-density wave (CDW) state, a structure in the self-
energy is detected at $\sim 65$ meV that cannot be explained by electron-phonon 
scattering. A reduction in the scattering rates below this energy indicates the 
collapse of a major scattering channel with the formation of the CDW state 
accompanying the appearance of a bosonic "mode" in the excitation spectrum of 
the system.
\end{quote}}}
\maketitle
In general the quasi 2-D electronic systems have a weaker tendency towards the 
formation of Charge-Density-Wave (CDW) and Spin-Density-Wave (SDW) instabilities 
than their 1-D counterparts. This is because the Fermi surfaces in 2-D 
(generalized cylinders) can only be partially nested. However, under favorable 
nesting conditions, or driven by saddle-point singularities, the electronic 
susceptibility may be enhanced enough for a CDW to develop. Since the Fermi 
surface is only partially gapped, the system may retain metallic character even 
in the CDW state. The 2-D character and the existence of an anisotropic gap make 
these systems similar to the high $T_C$ superconductors (HTSCs). In particular, 
the low-energy one-electron-like excitations ("quasiparticles") may show certain 
similarities. Angle resolved photoelectron spectroscopy (ARPES) represents a 
powerful technique for studying the one-electron properties in low-dimensional 
materials. It measures the occupied component of the
single-particle spectral function A$({\bf k},\omega)$, thus providing direct 
insight into the fundamental many-body interactions of the system. The technique 
has the important advantage that it momentum resolves. Studies of momentum-
resolved gaps and self-energies, as well as their temperature dependencies, can 
help in understanding the fundamental mechanisms responsible for producing a 
variety of phases in strongly correlated low-dimensional systems. As such, not 
only are the more exotic systems currently under investigation, but many 
"simple" systems are also being reinvestigated. Indeed, in recent studies, the 
effects of the electron-phonon coupling on single-particle states have been 
directly observed \cite{1,2}. 
The change in a state's width and dispersion near the Fermi level provide a 
signature of the electron-phonon coupling. However, similar effects have been 
detected in some systems where such a correspondence cannot be established. 
For example, in the cuprate superconductor, Bi$_2$Sr$_2$CaCu$_2$O$_{8+\delta}$
, a "kink" in the dispersion 
and a narrowing of the state, has been also detected \cite{3}. However, in that 
system, the doping and temperature dependencies rule out the electron-
phonon coupling as the mechanism and point towards a different type of coupling 
resulting in modifications of the single-particle spectrum \cite{4}. 

In the present paper, we report a detailed ARPES study of low-energy electronic 
excitations in 2H-TaSe$_2$. 2H-TaSe$_2$ undergoes a second-order transition to 
an incommensurate CDW at 122 K, followed by a first-order lock-in transition to 
a $3\times3$ commensurate CDW phase at 90 K \cite{5}. The driving mechanism for 
the CDW transition is still under debate. The two most recent ARPES studies have 
provided alternative explanations: Liu {\it et al} \cite{6} found the opening of 
a large gap in the extended saddle band region when the system entered the CDW 
state; Straub {\it et al} \cite{7}, in a study of 2H-NbSe$_2$, concentrated on 
the nesting properties of the Fermi surface, and found large portions that, 
although not gapped, could be nested. 
In both studies, the corresponding "nesting vectors" cannot be easily related 
to the CDW wave vector. 

In the present study we concentrate on the properties of the one-particle 
excitation spectrum and provide evidence of a possible collective mode that 
exists in the CDW state. This mode causes significant re-normalization of the 
one-particle spectrum at low energies. 
Simultaneously, the one-particle scattering rates at low energies are greatly 
reduced in the CDW state, indicating a collapse of the phase space available for 
scattering. The mode may be attributed to the electron-hole pair creation in 
those regions that are gapped, or more precisely, to the scattering from fluctuations associated with the CDW order parameter. The observation of a mass 
or velocity 
renormalization in the ungapped region of the Fermi surface accompanying the 
formation of a gap elsewhere in the zone is very reminiscent of the behavior 
observed for the high Tc superconducting materials, as discussed above 
\cite{3,4}.

The experiment reported here was carried out on a high resolution photoemission 
facility based on undulator beamline U13UB at the National Synchrotron Light 
Source. This facility employs a Scienta SES-200 electron 
spectrometer which simultaneously collects a large energy (0.5 to 1 eV) and 
angular window $(\sim12^{\circ})$ of the photoelectrons. This reduces the time 
needed for data acquisition and ensures that the whole region of interest in 
$k$-space, is recorded under identical conditions of temperature and surface 
cleanliness. The combined instrumental energy resolution was set to $\sim6$ meV, 
small enough to make no significant contribution to the photoemission peak 
widths measured here. The angular resolution was better than $\sim0.2^\circ$, translating into a momentum resolution of $\sim0.005$ \AA$^{-1}$ at the 
15.2 eV photon energy used in the study. 
Samples, grown by a chemical reaction with iodine as a transport agent \cite{8}, 
were mounted on a liquid He cryostat and cleaved in situ in the UHV chamber with 
base pressure $3\times 10^{-9}$ Pa. The temperature was measured using a silicon 
sensor 
mounted near the sample and was rechecked by fitting the Fermi edge, taking 
account of the full experimental energy resolution. 
\begin{figure}
\centerline{\epsfxsize=9cm\epsfbox{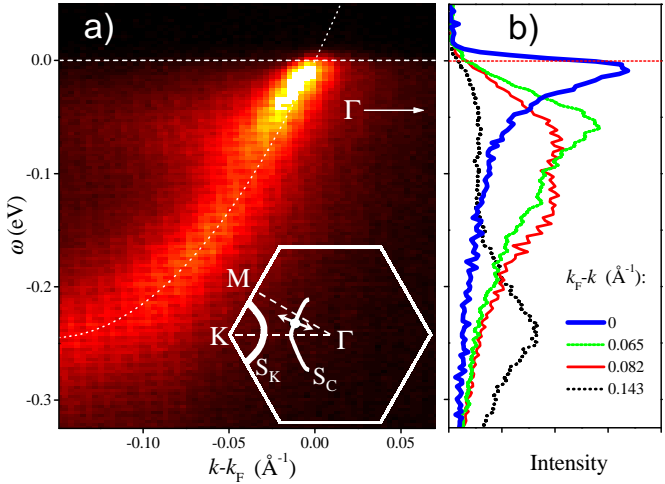}}
\caption{
(a) Photoemission intensity in the CDW state ($T=34$ K) as a function of binding energy and momentum along 
the line indicated in the inset by the double-headed arrow. Intensity is 
represented by a false color map, with yellow and white representing the highest 
intensity. The dispersing state is a part of the hole-like Fermi surface $S_C$, 
centered at $\Gamma$. This Fermi surface is not gapped in the CDW state. 
(b) EDCs, measured for several momenta as discussed in the text.}
\label{fig:1}
\end{figure}

Figure \ref{fig:1} shows the photoemission intensity, recorded in the CDW state 
at $T=34$ K, 
as a function of binding energy and momentum along the line through the two-
dimensional Brillouin zone indicated in the inset of the figure. The figure 
shows a band crossing the Fermi level at a point on the hole-like Fermi surface 
$S_C$, centered at $\Gamma$ \cite{double}. This particular Fermi surface is 
preserved in the 
CDW 
transition with no gap forming, independent of ${\bf k}_F$ \cite{6,7}. The most 
remarkable 
feature in figure \ref{fig:1} is the "kink" in the band's dispersion, 
accompanied by a sharpening in the vicinity of the Fermi level. Also 
shown in the figure are cuts through the intensity at constant momenta or energy distribution curves (EDCs). In this energy range, the EDCs show a two-peaked 
structure, behavior that is characteristic of the interaction of the photo-hole 
with some excitation of the system with energy range limited approximately to 
the energy scale of the "kink" (see \cite{1,2} and references therein). Such an 
interaction re-normalizes the mass and 
lifetime of the photo-hole but conserves the total charge. The self-energy, 
$\Sigma({\bf k},\omega)$, describes this interaction, the real part 
corresponding to the shift in 
energy and the imaginary part the scattering rate or inverse lifetime. Both 
components of the self energy may be extracted directly from an ARPES spectrum 
since the spectral intensity I$({\bf k},\omega)$ is given by I$^{0}({\bf 
k})$A$({\bf k},\omega)f(\omega)$ where A$({\bf k},\omega)$ 
represents the spectral function, I$^{0}({\bf k})$ incorporates the dipole 
matrix elements, 
and $f(\omega)$ is the Fermi distribution function. As recently noted by LaShell 
{\it et 
al} \cite{2}, in the limit of a momentum independent self energy and matrix 
elements, 
the spectral intensity takes the simple form:
\begin{figure}
\centerline{\epsfxsize=8.1cm\epsfbox{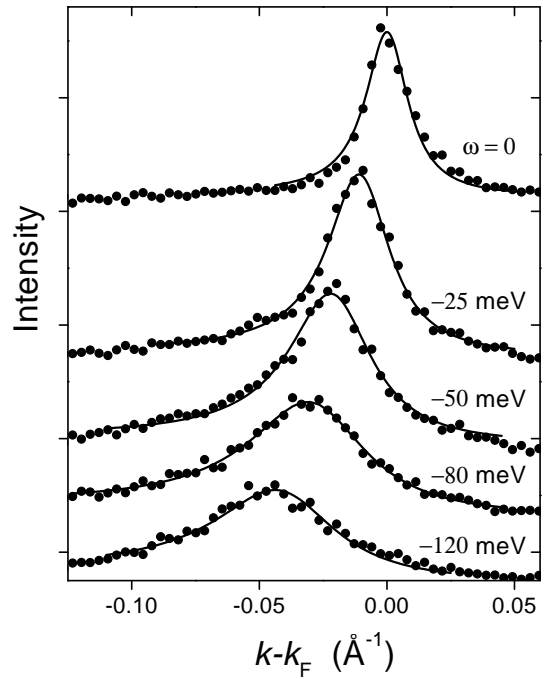}}
\caption{
MDCs, measured at different binding energies (symbols), fitted with a momentum-
independent spectral function (solid lines) as discussed in the text.}
\label{fig:2}
\end{figure}
\begin{equation}
\mathrm{I}(k,\omega)\propto \frac{{\mathrm Im}\Sigma(\omega)}{[\omega-
\epsilon_{{\bf k}}-\mathrm{Re}\Sigma(\omega)]^{2}+({\mathrm 
Im}\Sigma(\omega))^{2}}{\it f}(\omega)
\end{equation}
where $\epsilon_k$ is the non-interacting dispersion. The real and imaginary 
components of 
the self-energy, Re$\Sigma(\omega)$ and Im$\Sigma(\omega)$, may then be 
extracted directly from a 
momentum-distribution curve (MDC), the intensity as a function of momentum at 
constant binding energy. With this method, the fitting is possible without 
imposing any particular model for the interaction. We approximate the non-
interacting dispersion with a second-order polynomial \cite{9} that coincides 
with 
the measured dispersion at $k=k_F$ and at higher binding energies, close to the 
bottom of the band: Re$\Sigma=0$ for $\omega=0$ and $\omega<-200$ meV. Figure 
\ref{fig:2} shows several MDCs 
with corresponding fits. In contrast to the lineshapes in fig. \ref{fig:1}(b) 
for EDCs, 
the lineshapes in figure \ref{fig:2} are approximately Lorentzian at low binding 
energies 
developing an asymmetry at higher binding energies. The latter asymmetry mostly 
reflects the quadratic term in the non-interacting dispersion. The advantage of 
using MDCs in the analysis is obvious in that the self-energies are more 
dependent on energy than on momentum.

\begin{figure}
\centerline{\epsfxsize=9cm\epsfbox{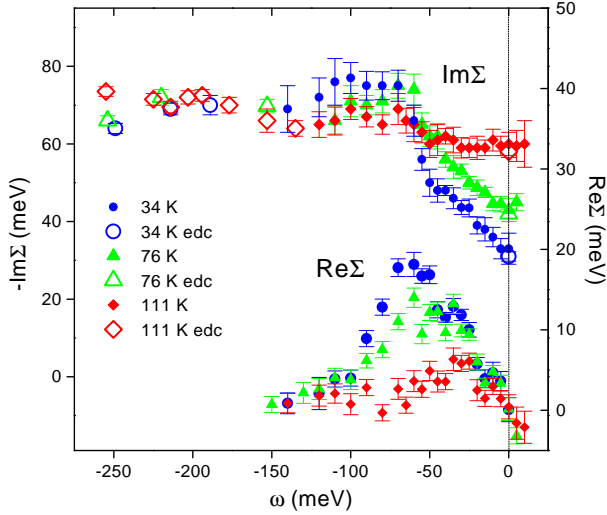}}
\caption{
Self-energy extracted from MDCs for several temperatures. Results for Im$\Sigma$ 
obtained from EDCs are shown as open symbols.}
\label{fig:3}
\end{figure}
The results of the fitting procedure, which produces pairs of Re$\Sigma$ and 
Im$\Sigma$ for 
every MDC are shown in Fig. \ref{fig:3} for several temperatures. We have also 
included 
Im$\Sigma$ obtained by fitting EDCs when the latter have a Lorentzian lineshape. 
The 
real part of the self-energy is concentrated in the region of binding energies 
less than 150 meV. At the lowest temperature, it has a maximum at a binding 
energy of $\sim65$ meV, approximately coincident with the value corresponding to 
the 
sharp drop in Im$\Sigma$. Such behavior is indicative of the scattering of 
the 
photo-hole from some collective excitation or "mode" of the system. The striking 
similarity with the behavior recently observed in an ARPES study of the photo-
hole interacting with phonons \cite{1,2} would point to the electron-phonon 
coupling 
as the source of this behavior. This would imply the presence of $\sim70$ meV 
phonons 
in the CDW state. However the highest calculated and measured phonon frequency 
is $\sim40$ meV \cite{10}. The measured temperature dependence of the self-
energy also 
argues against phonons. 
With increasing temperature, the peak in Re$\Sigma$ 
loses its 
magnitude and the structure shifts to lower energies. At a temperature of 111 
K, only a small peak is left at a binding energy of $\sim$30 meV and this 
survives in 
the normal state to at least 160 K. This peak may be of the same CDW origin, 
but may also be caused by conventional electron-phonon coupling, since it is 
within the range of the phonon spectrum.
\begin{figure}
\centerline{\epsfxsize=8.8cm\epsfbox{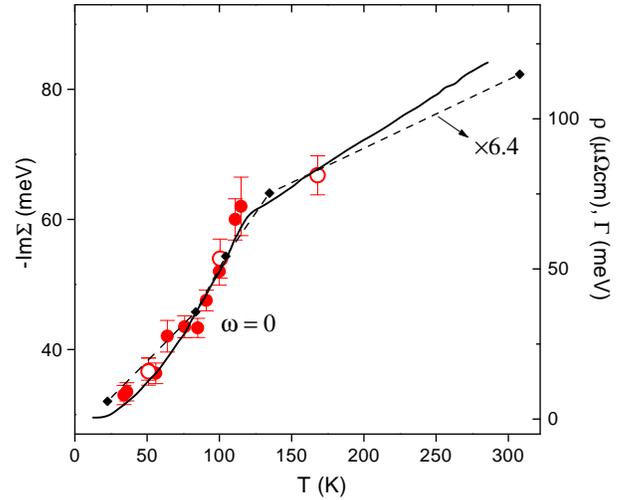}}
\caption{
Temperature dependence of the scattering rate at the Fermi level $(\omega=0)$. 
Resistivity $\rho(T)$ (solid line) and Drude scattering rate $\Gamma(T)$ (dashed 
line) from ref. [13] are given for comparison on the right scale.}
\label{fig:4}
\end{figure}

At low temperatures the imaginary part of the self-energy or scattering rate 
shows a sharp reduction for binding energies lower than 70 meV. As the 
temperature increases, this reduction becomes less pronounced. A more detailed 
temperature dependence for Im$\Sigma(0)$ is shown in Fig. \ref{fig:4}. The in-
plane resistivity 
and the Drude scattering rate measured by Vescoli {\it et al} \cite{11} are also 
shown. In a simple Drude-type model, the conductivity (the inverse of 
resistivity) in a 2-D system is proportional to the integral of $k_{F}l({\bf 
k}_{F})$ over the Fermi surface, weighted by geometrical factors defined by the 
field direction. Here $k_F$ is the Fermi wave vector and $l({\bf 
k}_{F})=1/\Delta k({\bf k}_{F})=v_{F}^{0}({\bf k}_{F})/{\rm 
Im}\Sigma(\omega=0,{\bf k}_{F})$ is the mean free path, $v_{F}^{0}({\bf k}_{F})$ 
being the bare Fermi velocity. 
The striking similarity between the scattering rates measured here and the 
resistivity indicates that this component of the Fermi surface plays an 
essential role in the transport. Indeed, similar 
behavior is found over the whole central Fermi surface, $S_C$, while the sections 
centered at K points and flat saddle regions show much higher scattering rates 
and/or are gapped in the CDW state \cite{6}. 
In addition to a significant zero-temperature 
offset \cite{12}, the single-particle scattering rate measured here has an 
approximately six times larger change over the same temperature interval than 
the Drude one. This is because transport currents are insensitive to small-
angle scattering, whereas ARPES is sensitive to all scattering events. 

The unusually large zero-temperature offset in the normal state resistivity
measurements may be caused by conventional impurity/defect scattering, which
usually adds a constant temperature-independent term to the normal state
resistivity. However, for 2H-TaSe$_2$ all published resistivity curves
\cite{11,13} appear to scale in a similar fashion over the whole temperature
range, indicating some form of intrinsic disorder in the normal state. As
suggested by Vecsoli {\it et al} \cite{11}, this may reflect the presence of
fluctuating incoherent CDW segments. To resolve this, it would be instructive
to study in more detail the effect of intentionally introduced defects on the
resistivity.

We have discussed the experimental observations in terms of scattering from some 
form of collective mode. It is also possible that the opening of the gap in 
the "saddle regions" with an associated reduction in the phase space available for 
scattering would be sufficient to explain the binding energy and temperature 
dependence of the self-energy. However, the opening of the gap in the single-
particle spectrum implies a modification to the response function with the 
possible simultaneous appearance of a new collective "mode". The self-energy 
behaves then as if the hole was scattered from this mode. The two pictures are 
equivalent if electronic correlations dominate. 

As pointed out by Vescoli {\it et al} \cite{11}, the in-plane optical response 
in 2H-TaSe$_2$ 
is very similar to that measured in high $T_C$ superconductors. In both 
systems one can resolve a Drude component and a mid-IR component in the gapped 
low temperature phases. As the Fermi surfaces in both systems are only 
partially gapped, it is tempting to connect the Drude component in the optical 
response to portions that are not gapped, while the mid-IR component 
seems to reflect the gapped regions. Whether the mid-IR structure itself points 
to the presence of the new mode is not clear. The presence of the well-known 
resonance in the spin response function of HTSCs shows that a mode with well-
defined quantum numbers may indeed be formed. Recently, the strong connection 
between the commensurate resonance magnetic peak in neutron scattering 
\cite{INS} and the mid-IR optical structure in HTSCs has been noted \cite{15}. 
Similarities in the optical response would then imply the 
possibility of a similar mode in the 2H-TaSe$_2$. More detailed studies of the 
charge and spin response of the system may help to resolve the nature of this 
mode.

In summary we have shown that the single-particle self-energy in 2H-TaSe$_2$ 
shows 
significant changes with the opening of the CDW gap implying the existence of a new 
type of collective excitation, associated with the CDW. Indeed, the observations 
suggest that the photo-hole is scattering from fluctuations in the CDW state. We 
believe that the 
observed behavior warrants further theoretical investigation in view of the 
strong similarities with the behavior observed in HTSCs. 
The authors would like to acknowledge useful discussions with L. Forr\'o, N. V. 
Muthukumar and A. Tsvelik. 
The BNL research was supported in part by the U.S. DOE under 
DE-AC02-98CH10886. The Boston University research was 
supported in part by the U.S. Department of Energy under 
DE-FG02-98ER45680. Experiments were undertaken at the NSLS, which is 
supported by the U.S. DOE, Divisions of Materials and Chemical Sciences.


\begin{references}
\bibitem{double} The Fermi surface of 2H-TaSe$_2$ is double walled, suggesting that 
every band should be split into two \cite{6}. Indeed, we have detected both bands 
and observed that the splitting, as well as the relative intensity of the two 
bands, is strongly dependent on momentum, photon energy, polarization and 
surface quality. In some circumstances, as in the case presented here, we obtained only one band. However, when both bands are well resolved, they show 
similar behavior in vicinity of the Fermi level and similar self-energy 
corrections to those discussed in the present study. Therefore the presence 
(or absence) of the second band is irrelevant here.
\bibitem{9} Note that an MDC is a Lorentzian centered at $k=k_{F}+[\omega-
{\mathrm Re}\Sigma(\omega)]/v^{0}$ with a width 
$\Gamma=2|{\rm Im}\Sigma(\omega)|/v^{0}$ only in the case of a linear bare 
dispersion, $\epsilon_{k}=v^{0}(k-k_{F})$.
\bibitem{12} Constant term measured in ARPES depends on surface quality and in 
the best cases was close to the limit of the instrumental resolution.
\end{references}
\end{document}